\documentclass[published]{JHEP3} 

\JHEP{00(2012)000}

\JHEPspecialurl{http://jhep.sissa.it/JOURNAL/JHEP3.tar.gz}

\usepackage{epsfig,multicol,bbm}

\newcommand\fverb{\setbox\fverbbox=\hbox\bgroup\verb}
\newcommand\fverbdo{\egroup\medskip\noindent%
			\fbox{\unhbox\fverbbox}\ }
\newcommand\fverbit{\egroup\item[\fbox{\unhbox\fverbbox}]}
\newbox\fverbbox

\title{Dark energy from dark radiation in strongly coupled cosmologies
with no fine tuning.}

\author{Silvio A. Bonometto$^{1,2,3}$, Giandomenico Sassi$^{4}$,
  Giuseppe La Vacca$^{4,5}$ \\ $^1$ -- Department of Physics,
  Astronomy Unit, Trieste University, Via Tiepolo 11, I~34143 Trieste,
  Italy \\ $^2$ -- I.N.F.N. -- Sezione di Trieste, Via Valerio, 2
  I~34127 Trieste, Italy \\ $^3$ -- I.N.A.F. -- Astronomical
  Observatory of Trieste, Via Tiepolo 11, I~34143 Trieste, Italy
  \\ $^4$ -- Departement of Physics G.~Occhialini, Milano--Bicocca
  University, Piazza della Scienza 3, 20126 Milano, Italy \\ $^5$ --
  I.N.F.N. -- Sezione di Milano Bicocca, Piazza della Scienza 3, 20126
  Milano, Italy\\ }

\received{.....}
\accepted{.....}

\abstract{A dual component made of non--relativistic particles and a
  scalar field, exchanging energy, naturally falls onto an attractor
  solution, making them a (sub)dominant part of the cosmic energy
  during the radiation dominated era, provided that the constant
  $\beta$, measuring the coupling, is strong enough. The density
  parameters of both components are then constant, as they expand as
  $a^{-4}$. If the field energy is then prevalently kinetic, as is
  expected, its energy is exactly half of the pressureless component;
  the dual component as a whole, then, has a density parameter
  $\Omega_{cd} = 3/4\beta^2$ (e.g., for $\beta \simeq 2.5$,
  $\Omega_{cd} \simeq 0.1$, in accordance with Dark Radiation
  expectations). The stationary evolution can only be broken by the
  rising of other component(s), expanding as $a^{-3}$. In a realistic
  scenario, this happens when $z \sim 3$--$5 \times 10^3$. When such
  extra component(s) become(s) dominant, the densities of the dual
  components also rise above radiation. The scalar field behavior can
  be easily tuned to fit Dark Energy data, while the coupled DM
  density parameter becomes $\cal O$$(10^{-3})$. This model however
  requires that, at present, two different DM components exist. The
  one responsible for the break of the stationary regime could be
  made, e.g., by thermally distributed particles with mass even $\gg
  1$--2 keV (or non--thermal particles with analogous average speed)
  so accounting for the size of observed galactic cores; in fact, a
  fair amount of small scale objects is however produced by
  fluctuation re--generated by the coupled DM component, in spite of
  its small density parameter, after the warm component has become
  non--relativistic.  }


\begin{document}

\section{Introduction}
Cosmological models supported by data are affected by a number of fine
tuning and coincidence paradoxes. Among them, we have the problems
concerning Dark Energy (DE), namely if its state parameter $w \equiv
-1 :$ (i) Why are we living in the only era when DE and matter have
comparable densities? (ii) Why density inhomogeneities are normalized
so that DE allows fluctuations to reach the non--linear regime, then
stopping any further density evolution?

Models trying to ease these paradoxes were proposed in the last few
years, often based on the idea that DE is a scalar field \cite{RP,SU,
  others}, possibly coupled to Dark Matter (DM) \cite{amendola}.
Current data, however, do not provide sufficient discrimination, so
that no such model really appears statistically favored in respect to
$\Lambda$CDM (see, e.g., \cite{colom}). Accordingly, rather than
suggesting models, recent work has focused on planning measures
allowing us to discriminate between a DE state equation $w(a) \equiv
-1$ and other behaviors, as the forthcoming Euclid
mission\footnote{http://www.euclid-ec.org} \cite{Laureijs:2011mu}

In this paper, in a sense, we partially go back to the older approach,
by suggesting a class of cosmological models, based on the assumption
that DE is a scalar field $\phi$. In describing them, however, no
peculiar self--interaction potential $V(\phi)$ is selected, as the
features we point out are (almost) independent from it, so that
changing $V(\phi)$ has a modest impact on our findings.  Accordingly,
while we expect that the shape of $V(\phi)$ can be hardly
discriminated from $\Lambda$CDM through CMB and related data fits
(see, e.g., \cite{wmap7}), {we suggest specific predictions possibly
  discriminating this class of models from different cosmologies.

The basic point made in this paper is that a large fraction (typically
1/3) of Dark Radiation (DR) is a scalar field $\phi$, later turning
into Dark Energy. Big--Bang Nucleosynthesis (BBN) is consistent with
observable nuclide abundances in the presence of an extra radiative
component consistent with $\Delta N_{off} < 1.26$ neutrino species
\cite{DRBBN}, and this constraint can be furtherly softened in the
presence of primeval lepton anti--lepton asymmetry. CMB data are
slightly more constraining, as the best fit of WMAP and related data
yields $\Delta N_{off} = 0.85 \pm 0.62$ (2 standard deviations)
\cite{DRCMB1}. A large deal of work has therefore been dedicated to
the nature of DR, also outlining that the amount of DR could be quite
different in BBN and CMB data \cite{DRCMB2}.

The price to pay for DR turning into DE is that two kinds of Dark
Matter (DM) must exist, that we shall dub $DM_{cou}$ and
$DM_{unc}$. The former one ($DM_{cou}$) is coupled to the scalar field
$\phi$. In the radiative era it is the rest of DR; in the Newtonian
limit, it has ordinary gravitational interactions with any other
cosmic components, but its fluctuations feel a much stronger self
gravity, while its dynamics has further specific modifications
\cite{simulations}. Its density parameter, in the present epoch, shall
lay in the per mil range, but its peculiar gravitational behavior can
give it an important role in shaping a large deal of today's
observables. The latter DM kind ($DM_{unc}$), expected to have an
ordinary gravitational behavior, is then needed to break a primeval
stationary condition. In principle, quite a few discriminatory
predictions may follow from the presence of the $DM_{cou}$ component,
coupled to DE.

Although the mechanism turning DR into DE is the basic point of this
work, we shall devote a specific Section to discuss possible scenarios
allowed by the presence of 2 DM kinds. Rather than a complication,
this appears as an ``opportunity''. In particular, we shall discuss
the case of $DM_{unc}$ being warm; the simultaneous presence of
$DM_{cou}$ might then ease a number of cosmological problems. Among
them, the small halo deficit and the size of the plateau in galaxy
cores, when comparing observations to simulations.

All that is obtained by pushing to extreme consequences the idea of
DE--DM coupling, as previously suggested in \cite{amendola}.  The
coupling intensity considered is however much greater than any
previous analysis, and consistency with data can be recovered thanks
to the fact that $DM_{cou}$ is a minor component of DM.

The plan of the paper is as follows: In the next Section we shall
discuss how to treat coupled DM and DE if the DE state equation $w(a)$
is assigned, while the DE self--interaction potential $V(\phi)$ is
unknown. The equation of motion to be solved is then simpler than the
usual Klein--Gordon equation, as the problem becomes first order, the
real unknown being $\phi_1 \equiv \dot \phi,$ while we do not need to
know $\phi$. In Section 3, we then consider the $\phi_1$ equation and
the equation ruling $DM_{cou}$ density in the radiative era, finding a
self--consistent solution, which allows constant density parameters
for $DM_{cou}$ and $\phi$--field, provided that the coupling strength
is large enough. In Section 4 we verify this solution to be an
attractor, and that we converge on it if starting from any initial
condition. In Section 5 we briefly discuss the form of the effective
Lagrangian yielding the equation of motion. Section 6 is then devoted
to debating what happens if another non--relativistic component
finalizes the radiative expansion. In Section 7 we then show that,
rather than a complication, the two DM components are an opportunity,
in particular if we assume $DM_{unc}$ to be WDM. A Discussion Section
concludes the paper.

\rm

\section{CDM--DE coupling}
The possibility that CDM and DE are coupled have been considered by
several authors \cite{amendola,spain,mmc}. As a matter of fact, while
the stress--energy tensors of CDM and DE, ${T_{(c)}}_{\mu\nu}$ and
${T_{(d)}}_{\mu\nu}$ respectively, surely fulfill the
pseudo--conservation equation
\begin{equation}
\label{eq1}
{T_{(c)}}^\nu_{\mu; \nu} + {T_{(d)}}^\nu_{\mu; \nu} = 0~,
\end{equation}
there is no direct evidence that the two equations
\begin{equation}
{T_{(c)}}^\nu_{\mu; \nu} = 0,~~~~~~ {T_{(d)}}^\nu_{\mu; \nu} = 0~
\end{equation}
are separately satisfied. The r.h.s. of the above equations can then
be replaced, in a covariant way, by a term yielding a leakage of
energy from DE to CDM or viceversa.

Here we shall consider the option yielding
\begin{equation}
\label{coupled}
\ddot \phi + 2{\dot a \over a} \dot \phi + a^2 V'_\phi(\phi) 
= C \rho_c a^2~,~~~~~
\dot \rho_c + 3{\dot a \over a} \rho_c = -C \dot \phi \rho_c~.
\end{equation}
Here $\rho_c$ is the density of CDM, $\phi$ is a scalar field
self--interacting through the potential $V(\phi)$ and accounting for
DE, while
\begin{equation}
\label{C}
C = {b \over m_p} = \sqrt{16 \pi \over 3} {\beta \over m_p} > 0
\end{equation}
accounts for energy transfer from CDM to DE. Taking a constant $\beta$
is an extra assumption we shall make here for simplicity. Here we also
assume a FRW metric reading
\begin{equation}
ds^2 = a^2(\tau)(d\tau^2 - d\ell^2)~,
\label{FRW}
\end{equation}
so that $a$ is the scale factor, $d\ell$ is the spatial line element,
while differentiations are made in respect to the conformal time
$\tau$.

The option (\ref{C}) has a peculiar significance, as it allows models
where DE is always a significant component of the Universe, being fed
energy by the CDM component which, accordingly, dilutes more rapidly
than $a^{-3}$.

Let us then reconsider the eqs.~(\ref{coupled}) when the potential
$V(\phi)$ is unknown, while we know that the DE state equation is
a suitable $w(a)$. It must however be
\begin{equation}
\label{potential}
w = { \dot \phi^2/2a^2 - V \over \dot \phi^2/2a^2 + V }~,
~~~~{\rm i.e.}~~~~~
V = {\dot \phi^2 \over 2} {1-w \over a^2(1+w)}~,
\end{equation}
and $a^2V'_\phi$, in eq.~(\ref{coupled}), can be directly evaluated by
considering how the two factors in the last expression depend on
$\tau$, then associating such dependence with the $\tau$ dependence
of~$\phi$.

Let us then notice that:
\begin{equation}
 {d \over d \phi} {\dot \phi^2  \over2} =  
 {d \over d\dot \phi} {\dot \phi^2 \over2} 
\times {d \dot \phi
\over d\tau} \times {d \tau \over d\phi} = \ddot \phi~,
\end{equation}
\begin{equation}
 {d \over d a}  \left({1 \over a^2} {1-w \over 1+w}\right) =
-  {2 \over a^3}  \left[{1-w \over 1+w} + {dw \over da} {a \over (1+w)^2}
\right]
\end{equation}
so that 
\begin{equation}
a^2 V'_\phi = \ddot \phi {1-w \over 1+w} - {\dot a \over a}\dot \phi 
\left[ {1-w \over 1+w} + {dw \over da} {a \over (1+w)^2} \right]~.
\end{equation}
Let then
\begin{equation}
\tilde W = {1 \over 2}\left(1+3w-{a \over 1+w}{dw \over da}\right)
\end{equation}
in order that the system of equations (\ref{coupled}) becomes
\begin{equation}
\label{c1}
\dot \phi_1 + \tilde W{\dot a \over a} \phi_1
= {1+w \over 2} C \rho_c a^2~,~~~~~
\dot \rho_c + 3{\dot a \over a} \rho_c = -C \phi_1 \rho_c~.
\end{equation}
Here we set $\phi_1 = \dot \phi$ to outline that the former equation
has become first order. 

As a matter of fact, it seems more realistic that data allows us to
know $w(a)$, rather than the potential $V(\phi)$. Should we know
$w(a)$ and wish to interpret the $w(a)$ dependence as due to the
evolution of a scalar field, we need to integrate just
eqs.~(\ref{c1}), together with the Friedman equation. The scalar field
contribution to its source term is then the total energy density for
DE,
\begin{equation}
\label{total}
\rho_d = {\phi_1^2 \over a^2(1+w)}~,
\end{equation}
as also the potential contribution is derived from $\phi_1$ and
$w(a)$, thought eq.~(\ref{potential}). Apparently, therefore, we
need not recovering the un--differentiated $\phi$ behavior.

This however assumes that we know DE and $DM_{cou}$ to be coupled,
that the coupling is constant, and the coupling constant has a
specific value $C = 4(\pi/3)^{1/2}\beta/m_p$.

It is premature to discuss here observational strategies. A natural
start point, however, amounts to assuming no coupling. The apparent DE
state parameter, measured from from the expansion rate, would then be
\begin{equation}
\label{213}
w_{eff}(a) = w(a)/[1+\xi(a)]
\end{equation}
with
\begin{equation}
\xi(a) =  [g(\phi)/g(\phi_0)-1] \times \rho_{0c}/(\rho_d a^3)~.
\end{equation}
Here $g(\phi) \propto \rho_c a^3$ tells us the deviation of $\rho_c$
from the ordinary $a^{-3}$ scaling; the suffix $_0$ refers anywhere to
today's quantities (see \cite{cora} for a detailed discussion).

Notice that, in order to pass from $w(a)$ to $w_{eff}(a)$ or
viceversa, we then need to know $\phi$, besides of $\phi_1$.  However,
to recover $\phi$, we do not need integrating an equation, but just a
known function. 

We shall not delve here into a possible more refined analysis, seeking
the family of $w_\beta(a)$ behaviors as the assumed coupling $\beta$
varies. 

Notice also that eq.~(\ref{total}) shows that $\rho_d$ is positive
definite only if $w > -1$: to explore the $w < -1$ domain one needs
the $\phi$ field to have a suitable anomalous kinetic energy
expression.

These equations, therefore, require no specific potential shape to be
assigned, no background expansion regime to be assumed, while the very
$w(a)$ behavior is generic, although it must be $w(a) > -1$.

\section{Coupled DE in the radiative era}
Let us now consider the system (\ref{c1}) when the background
expansion is supposed to be radiative. We shall make the further
assumption that
\begin{equation}
\tilde W = {1 \over 2}(1+3w)~,
\end{equation}
with $w = {\rm const.}$, as is reasonable when $a \to 0$. This
assumption is however unessential and only allows us to simplify the
analytical treatment.

Let us also remind that, in the radiative era, it is $a \propto \tau$,
so that $\dot a/a = 1/\tau$. Accordingly, from Friedmann equations
we obtain that
\begin{equation}
{8 \pi \over 3 m_p^2}\,  \rho \, a^2 \tau^2 = 1~,
\label{frie}
\end{equation}
$\rho$ being the background energy density and $m_p$ the Planck mass.
The latter eq.~(\ref{c1}) has then the formal integral
\begin{equation}
\rho_c = \rho_{i,c} \left(a_i \over a \right)^3
\exp\left(-C\int_{\tau_i}^\tau d\tau \phi_1 \right)~,
\label{formal}
\end{equation}
$\tau_i$ being a reference time when CDM density is $\rho_{c,i}$ and
the scale factor is $a_i= a(\tau_i)$.  If this expression for $\rho_c$
is then replaced in the former eq.~(\ref{c1}), we have a first order
transcendental differential equation whose unknown is $\phi_1$. It
seems hard to find a generic analytic integral of this equation.

There is however a peculiar case, allowing integration. Let us
make the ansatz that
\begin{equation}
\label{phi1}
\phi_1 = \alpha {m_p \over \tau}~.
\end{equation}
Taking eq.~(2.4) into account, we have then that
\begin{equation}
-C\int_{\tau_i}^\tau d\tau \phi_1 
= \ln\left(\tau_i \over \tau \right)^{\alpha b}~,
\end{equation}
so that
\begin{equation}
\rho_c = \rho_{i,c} \left(a_i \over a \right)^{3+\alpha b}
\label{informal}
\end{equation}
and, by replacing the expression (\ref{phi1}) in the former
eq.~(\ref{c1}), we obtain
\begin{equation}
(\tilde W -1) \alpha {m_p \over a^2 \tau^2} = {1+w \over 2} {b 
\over m_p} \rho_{rc} \left(a_r \over a \right)^{3+\alpha b}~.
\label{same}
\end{equation}
In order that the two sides scale with $a$ in the same way, it must
then be $\alpha b = 1$ and the DM density shall scale with $a^{-4}$.
The fact that $\rho_c$ dilutes more rapidly than $\propto a^{-3}$ does
not come as a surprise, as there is a continuous leakage of energy
from it to the $\phi$ field. The fact that it dilutes exactly as
$a^{-4}$, instead, is a consequence of the ansatz (\ref{phi1}).

Equation (\ref{same}) can then be put in the form
\begin{equation}
{1 \over \beta^2}{\tilde W -1 \over 1+w} 
=  {8 \pi \over 3 m_p^2}\,  \rho_c \, a^2 \tau^2
\equiv \Omega_c ~,
\end{equation}
owing to eq.~(\ref{frie}). Here $\Omega_c = \rho_c /\rho$ is the
(constant) density parameter of DM, during radiation era. In order
that the ansatz (\ref{phi1}) is allowed, $\Omega_c$ ought to have the
value given by this equation.

Also the energy density $\rho_d$ of the DE field $\phi$ scales with
$a^{-4}.$ In fact, owing to eq.~(\ref{total}),
\begin{equation}
\rho_d = {\alpha^2 m_p^2 \over a^2 \tau^2 }{1 \over 1+w}~,
\end{equation}
and, using again eq.~(\ref{frie}), we obtain the constant density
parameter of DE
\begin{equation}
\Omega_d = {1 \over 2\beta^2 (1+w)}
\end{equation}
showing also that
\begin{equation}
{\Omega_c \over \Omega_d} = 2(\tilde W -1 )= 3w-1~.
\end{equation}
Accordingly, the whole framework is consistent only if $w > 1/3$.
This means that the energy density of the kinetic part of the $\phi$
field should be dominant in respect to the potential part. In the
specific case $w \simeq 1$, holding for $\phi_1^2 \gg 2a^2V$, we
therefore expect that it is constantly
\begin{equation}
\Omega_c \simeq 2 ~\Omega_d
\end{equation}
during such expansion regime.

Altogether, these computations show that, during a radiative
expansion, we can have two coupled components also expanding $\propto
a^{-4}$ although their state equations are $w_c = 0$ and $w > 1/3$.
The former component is made of non--relativistic particles, the
latter is a self--interacting scalar field. A possible option is that
the individual CDM particle masses decrease in time, as a consequence
of their interaction with $\phi$. The two dark components interact
with a strength measured by the dimensionless parameter $\beta$.

Owing to their behavior, these two components do not modify the
radiative character of the expansion. However, we might prefer that
ordinary radiation is the dominant component during this period.  This
requires a strong coupling between the components $\beta \gg 1$, as it
should however be $w \leq 1~.$ In turn, the ratio between the density of
DM and DE is $\cal O$$(1)$. 

In the specific case $w \simeq 1$, however reasonable in the very
early Universe, we then have $\Omega_d \beta^2 = 1/4$ and
\begin{equation}
(\Omega_c+\Omega_d) \beta^2 = 3/4~.
\end{equation}
In an early epoch it is fair to assume a total density parameter
$\Omega_t = 1$. Then, requiring $\Omega_c + \Omega_d < 1$ yields
\begin{equation}
\beta > \sqrt{3}/2 = 0.866~.
\end{equation}
A solution with $w \simeq 1$ and $\beta^2 \simeq 3/4~$, although
self-consistent, seems unreasonable. In fact, then $\Omega_c+\Omega_d
\simeq 1$ and the ordinary radiation component should vanish.

Solution with $\beta^2 < 3/4$ require $w > 1$ to allow constant
$\Omega_{c,d}$. If $w > 1$ is excluded, when $\beta^2 < 3/4$ there
exist no solution with constant $\Omega_{c,d},$ i.e., the CDM and
$\phi$ field contributions to the overall density become increasingly
small when $a$ tends to zero.
\begin{figure}
\begin{center}
\vskip -.4truecm
\includegraphics[scale=0.44]{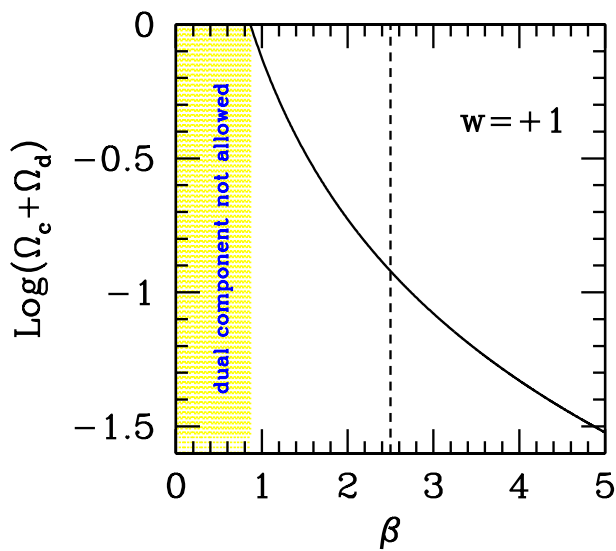}
\end{center}
\vskip -1.6truecm
\caption{Density parameter of the dual radiative component
vs.~the coupling parameter $\beta$ }
\label{odc}
\end{figure}

In Figure \ref{odc} we plot the density parameter of the dual (DM+DE)
component vs.~assumed $\beta$ values.

The dual component gives place to a natural form of DR. In general, we
can gauge its significance through the number of extra neutrino
species
\begin{equation}
\Delta N_{off} = {\rho_{DR} \over {\pi^2 \over 30} {7 \over 4} \left(
4 \over 11 \right)^{4/3} T^4}
\end{equation}
(here $\rho_{DR}$ is the dual--component density and $T$ is photon
temperature). It is then
\begin{equation}
\Delta N_{off} = {(8/7)(11/4)^{4/3}+3 \over (4/3)\beta^2  - 1}
 \simeq {7.4032 \over (4/3)\beta^2  - 1}
\end{equation}
and $\beta = 1.5$ (2.5) yields $\sim 3.7$ ($1.1$) extra species.

Let us recall that BBN prescribes $\Delta N_{off} < 1.26$, in standard
theories, or $\Delta N_{off} < 2.56$ when a non--vanishing chemical
potential is allowed for neutrinos \cite{DRBBN}. The overall density
of radiative components also sets the equality redshift. CMB and
related data analysis then allow us to fix the equality, so requiring
$\Delta N_{off} \simeq 0.85 \pm 0.62$ or similar figures
\cite{DRCMB1}, although the choice of priors risks to affect the final
estimate \cite{100}. All above limits are at $95\, \%$ confidence
level. Let us also recall that models can be easily built, where the
DR density at equality exceeds the one at BBN. The smallest coupling
consistent with all above limits is however around $\beta = 2~.$
Through this paper, only the case $\beta = 2.5$ will be however
considered.

\begin{figure}
\begin{center}
\vskip -.4truecm
\includegraphics[scale=0.37]{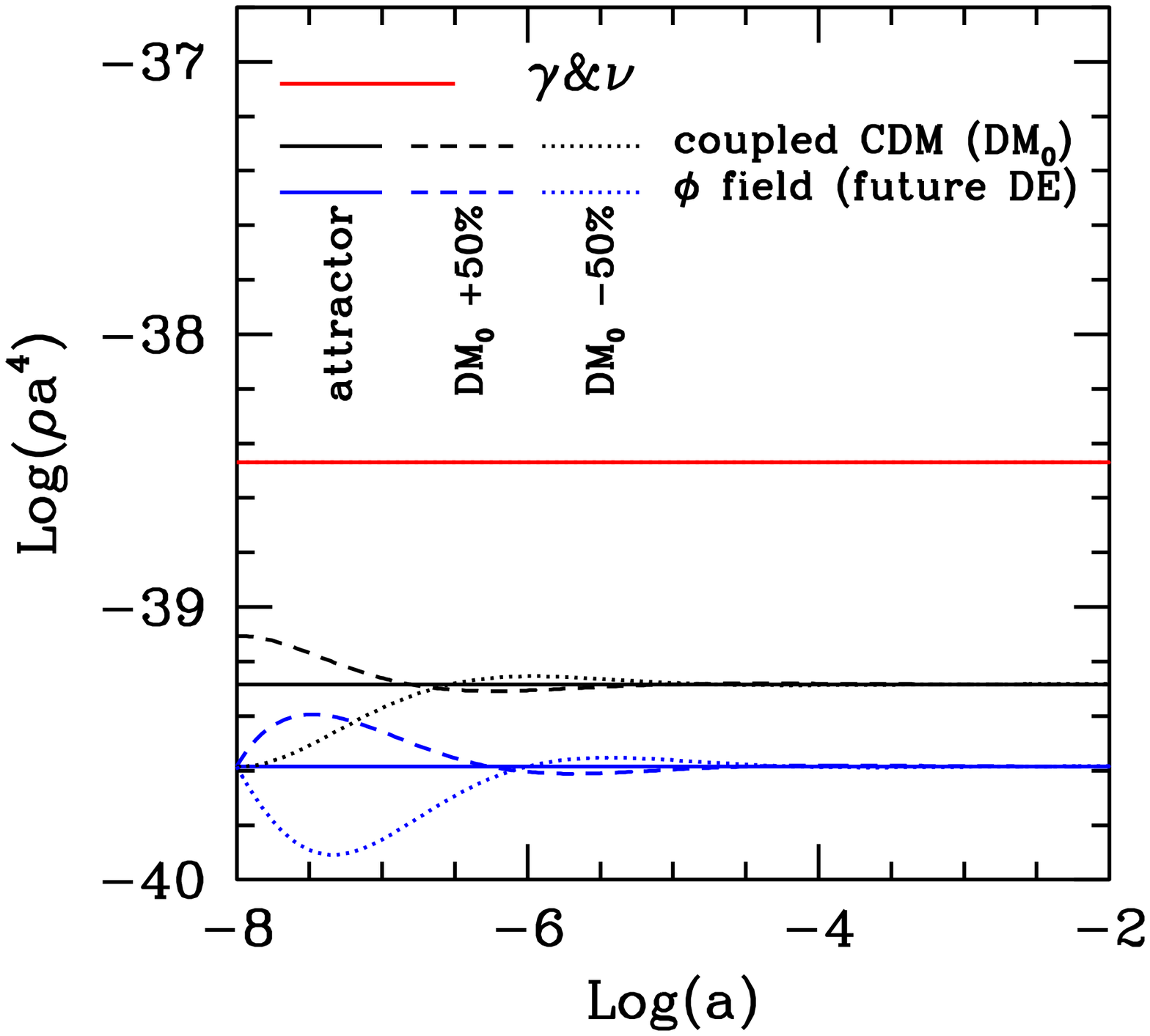}
\includegraphics[scale=0.37]{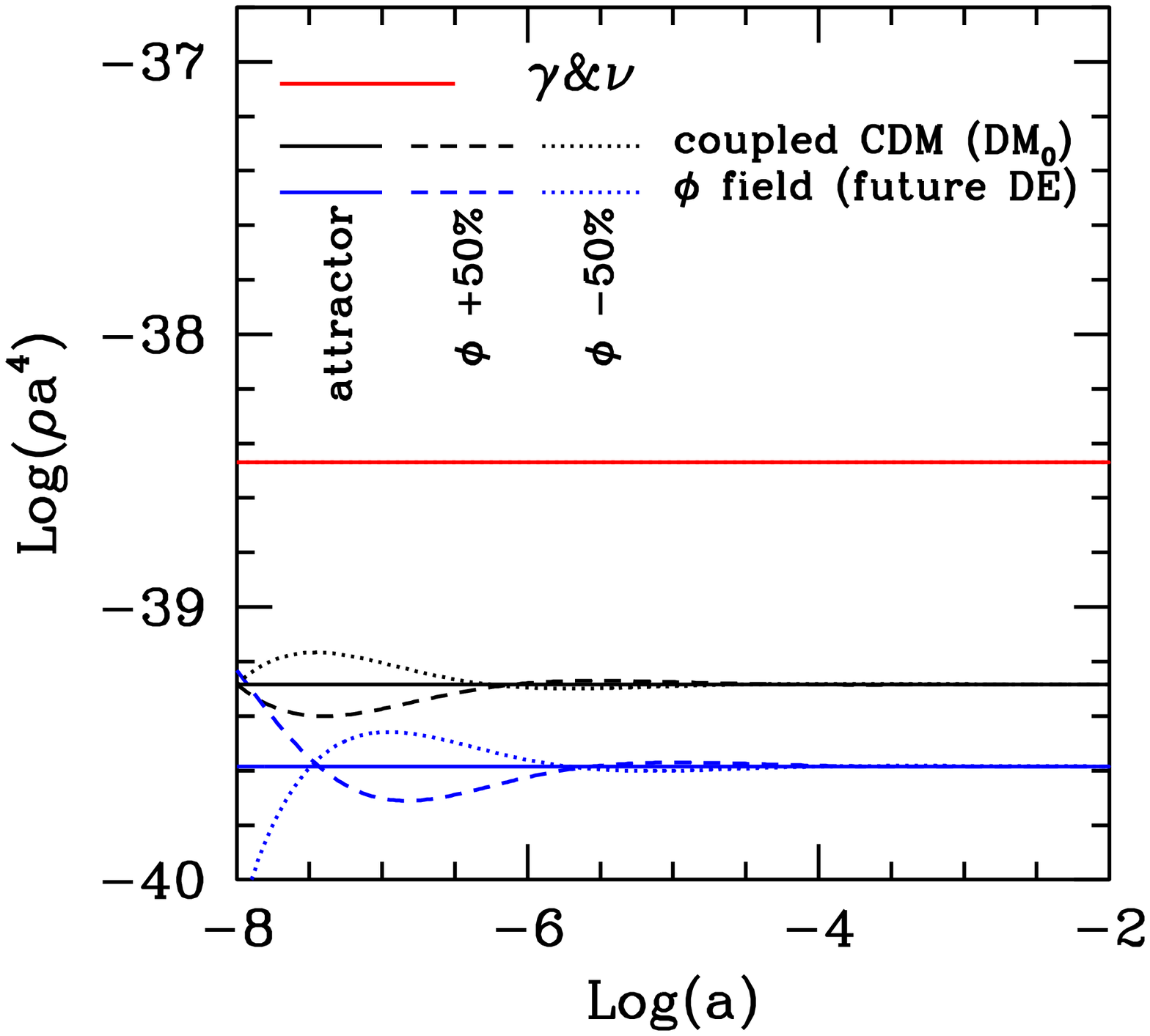}
\vskip -.6truecm
\includegraphics[scale=0.37]{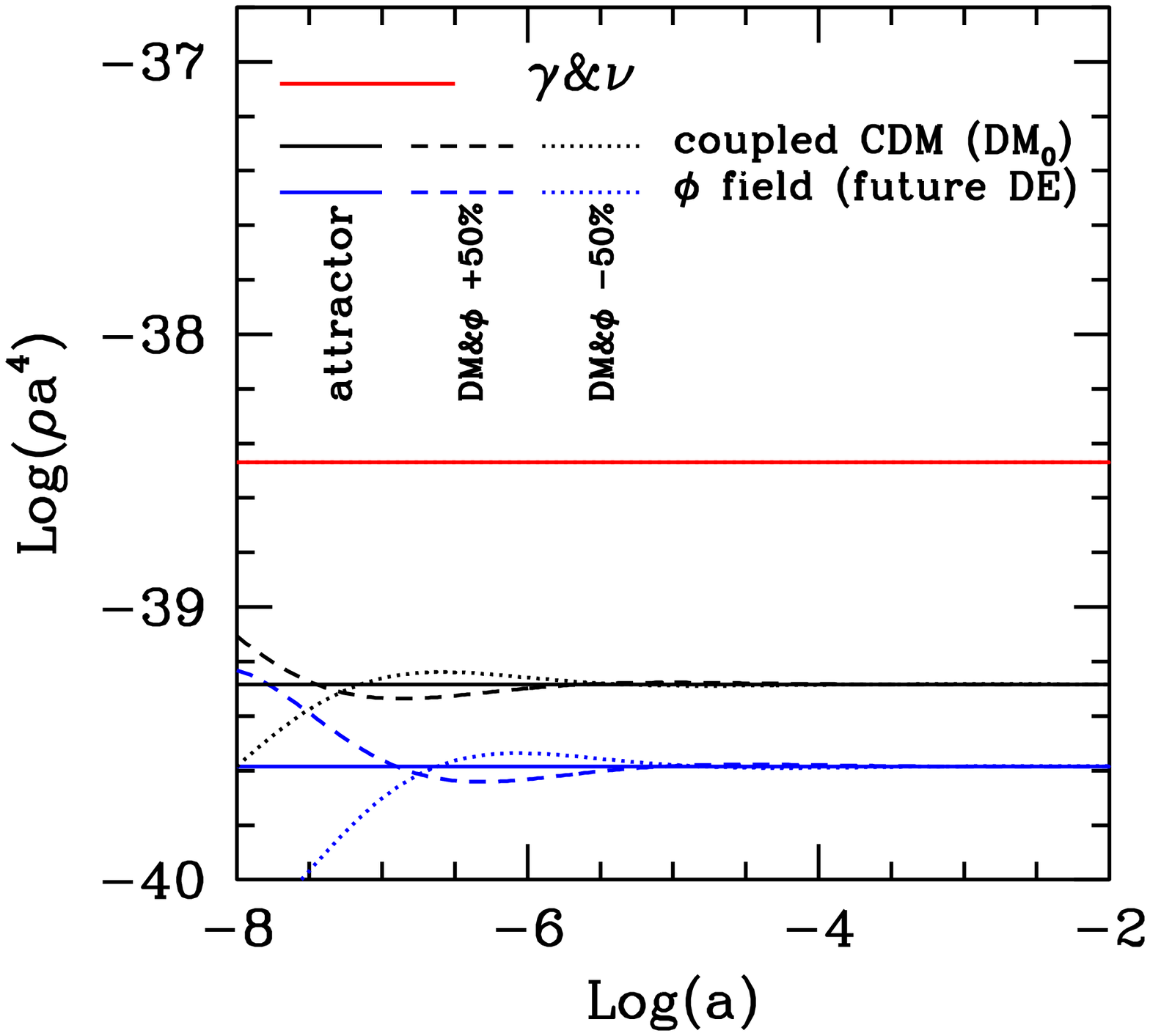}
\end{center}
\vskip -.7truecm
\caption{Stability of solutions. We start from modified initial
  values, showing that we soon reconverge on the ``attractor''
  solution. More specifically, we shifted by $\pm 50\, \%$,
  respectively: $\rho_c$, $\phi_1$, both of them; the last option is
  equivalent to setting a ``modified'' $\beta$ value. The solid lines
  are the ``attractor'' solution. Dashed and dotted lines show the
  gradual recovery of it. In these Figures $\beta = 2.5~.$ }
\label{attractor}
\end{figure}

\section{Stability}
Solutions with $w=1$ and
\begin{equation}
\label{attract}
 \Omega_c=2\, \Omega_d = 1/2\beta^2
\end{equation}
are however stable. If the initial values of $\Omega_c$, $\phi$, or
$\beta$ do not fulfil the above relation, and the expansion regime is
radiative, the condition (\ref{attract}) is soon restored.

In Figure \ref{attractor} we show this in 3 cases: (i) If we set an
initial value of $\rho_c$ in excess by 50$\, \%$. (ii) If we set an
initial value of $\phi_1$ in excess by 50$\, \%$. (iii) If both shifts
are simultaneously performed; this is equivalent to having a ``wrong''
initial $\beta$.

The results in Figure \ref{attractor} are obtained by numerically
integrating the set of differential equations (\ref{coupled})
plus the Friedmann equation
\begin{equation}
(\dot a/a)^2 = 8\pi \rho/3m_p^2~,
\label{fried}
\end{equation}
$\rho$ being the total background density.  All quantities are
expressed in MeV; in particular the units for the conformal time
$\tau$ (not shown) are MeV$^{-1}$.

The above numerical output can be analytically understood.  Let us
consider, e.g., that the density of the DE component has a value
different from what eq.~(3.10) requires, i.e. that
\begin{equation}
\phi_1 = \phi_{1,o} + \delta ~~~~~{\rm and}~~~~~ \phi_{1,o}=\alpha
m_p/\tau
\end{equation}
with $\delta $ positive or negative.  Assuming that $\rho_c$ is
unmodified, eq.~(\ref{c1}) tell us that $\delta$ must fulfill the
equation
\begin{equation}
\dot \delta + (1+\epsilon) \delta/\tau = 0
\end{equation}
with $\epsilon = \tilde W-1 > 1$ (for $w>1/3$). Then, $\delta \propto
\tau^{-(1+\epsilon)}$ and the ratio
\begin{equation}
\label{decre}
|\delta| / \phi_1 \propto \tau^{-\epsilon}
\end{equation}
necessarily decreases with time. As a matter of fact, however,
$\phi_1$ appears also in the differential equation setting $\dot
\rho_c$, while the impact on Friedmann equation is small if we assume
a large $\beta$ yielding small $\Omega_{c,d}$ (plots are for
$\beta=2.5$). A greater $\phi_1$ then yields a $\rho_c$ decrease
faster than $a^{-4}$. In turn this corresponds to a decreased energy
leakage towards the scalar field, so that $|\delta|$ decline is
accelerated, in respect to (\ref{decre}).  Figure \ref{attractor}
however shows that some bounces occur before the ``attractor''
solution is recovered.

Analogous arguments can be put forward for the cases (ii) and (iii).
For the sake of completeness, let us also outline that solutions are
tendentially stable also for constant $w < 1$. However, if initial
conditions violate eqs.~(3.10)--(3.11), the recovery of the attractor
solution takes an increasingly longer time as $w$ is farther from
unity.

\section{A Lagrangian approach}

The function $g(\phi)$ needed to define $\xi$ in eq.~(\ref{213}) also
enters in the Lagrangian coupling between the scalar $\phi$ field and
a supposed spinor field $\psi$, yielding $DM_{cou}$.

According to \cite{cora}, the general expression of the effective
interaction Lagrangian reads
\begin{equation}
{\cal L} = \mu g(\phi) \bar \psi \psi~,
\end{equation}
$\mu$ being a factor with the dimensions of a mass. Owing to
eq.~(\ref{formal}), then, it should be
\begin{equation}
g(\phi) = \exp\left( -b\int_{\tau_p}^\tau d\tau \phi_1/m_p \right)
\end{equation}
so that, owing to the ansatz (\ref{phi1}),
\begin{equation}
{\cal L} = \mu {\tau_p \over \tau} \bar \psi \psi = b {\mu \over m^2_p}
\dot \phi \bar \psi \psi = \Gamma \dot \phi \bar \psi \psi~,
\end{equation}
with a constant $\Gamma$ whose dimensions are $m^{-1}$, or
\begin{equation}
{\cal L} = {\mu\, T \over m_p} \bar \psi \psi~.
\end{equation}
This makes clear that the very interaction Lagrangian also displays
the role of variable--mass term, with a mass $\propto \tau^{-1}$ or
$\propto T$.

\rm

\section{Exit from the radiative regime}
\begin{figure}
\begin{center}
\vskip -.4truecm
\includegraphics[scale=0.50]{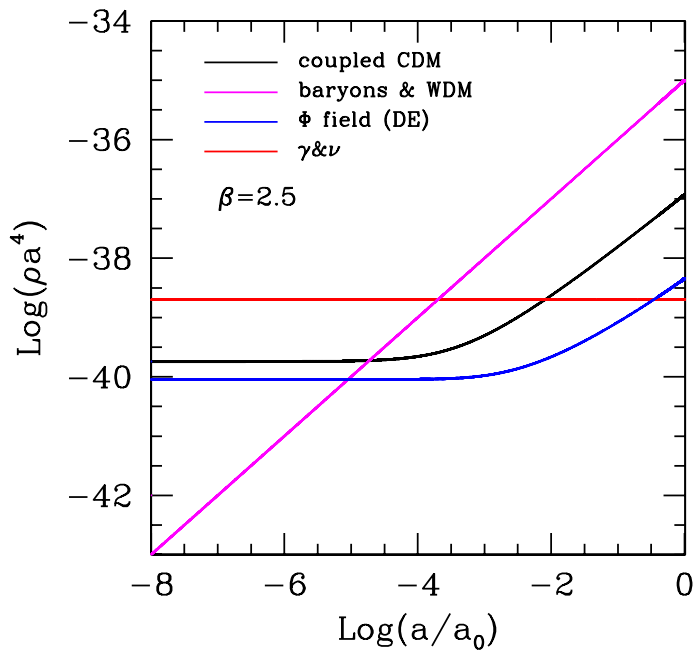}
\end{center}
\vskip -.6truecm
\caption{Density evolutions when a non--relativistic component becomes
  dominant at $z \simeq 5 \times 10^3$. Both DM$_{cou}$ and DE densities
  then overcome radiation, but keep steadily below DM$_{unc}$.  We assumed
  here $\beta=2.5$, so that $\Omega_c+\Omega_d \equiv 0.12$ in the
  radiative era; this corresponds to $\Delta N_{off} =1.009$, as Dark
  Radiation, i.e. to $\simeq $1 extra neutrino species.  }
\label{allc0}
\end{figure}
The picture changes if the expansion is no longer radiative.  In the
physical world, there will be baryons, at least, whose density
overcomes the density of the radiative components slightly below $z
\sim 10^3$. If their abundance is consistent with BBN, they are surely
not enough to produce an overall picture possibly close to
observations. We shall therefore assume that another non--relativistic
DM component exists so that its density summed to baryons matches the
density of the radiative component at $z = 5 \times 10^3~.$ We have
therefore two DM components: DM$_{cou}$ coupled to the field $\phi$;
DM$_{unc}$ uncoupled; for the sake of simplicity, here baryons are
included in DM$_{unc}$.

In Figure \ref{allc0} we show how DM$_{cou}$ and the DE field no longer
scale as radiation, when the overall expansion ceases to be radiative.
We assume that $\beta = 2.5$, so that $ \Delta N_{off} \simeq 1~, $
and that the equation of state of DE keeps $w = +1$ until today. Then,
$\phi$ does not contribute much to today's overall density while, by
decreasing $\beta$, we could have DM$_{unc}$ approaching the present
baryon density, at most. Below, we shall further comment on DM$_{unc}$
observable effects.

The situation changes radically if we include the option that the
$\phi$ field abandons the kinetic regime. Of course, this is necessary
if we wish to identify it with DE, whose today's state equations
approaches $w \simeq -1$. For most potentials considered in the
literature, when $\phi$ overcomes a suitable level, the potential
energy becomes indeed dominant. The moment when the transition occurs
depends on the potential assumed as well as on the initial
contribution of the two coupled components to the cosmic budget.

As an example, we report here the expected behavior of $w(z)$ when the
potential is SUGRA \cite{SU} or RP \cite{RP}. More details on these
plots are given in \cite{mmc}.
\begin{figure}
\begin{center}
\vskip -4.truecm
\includegraphics[scale=.52]{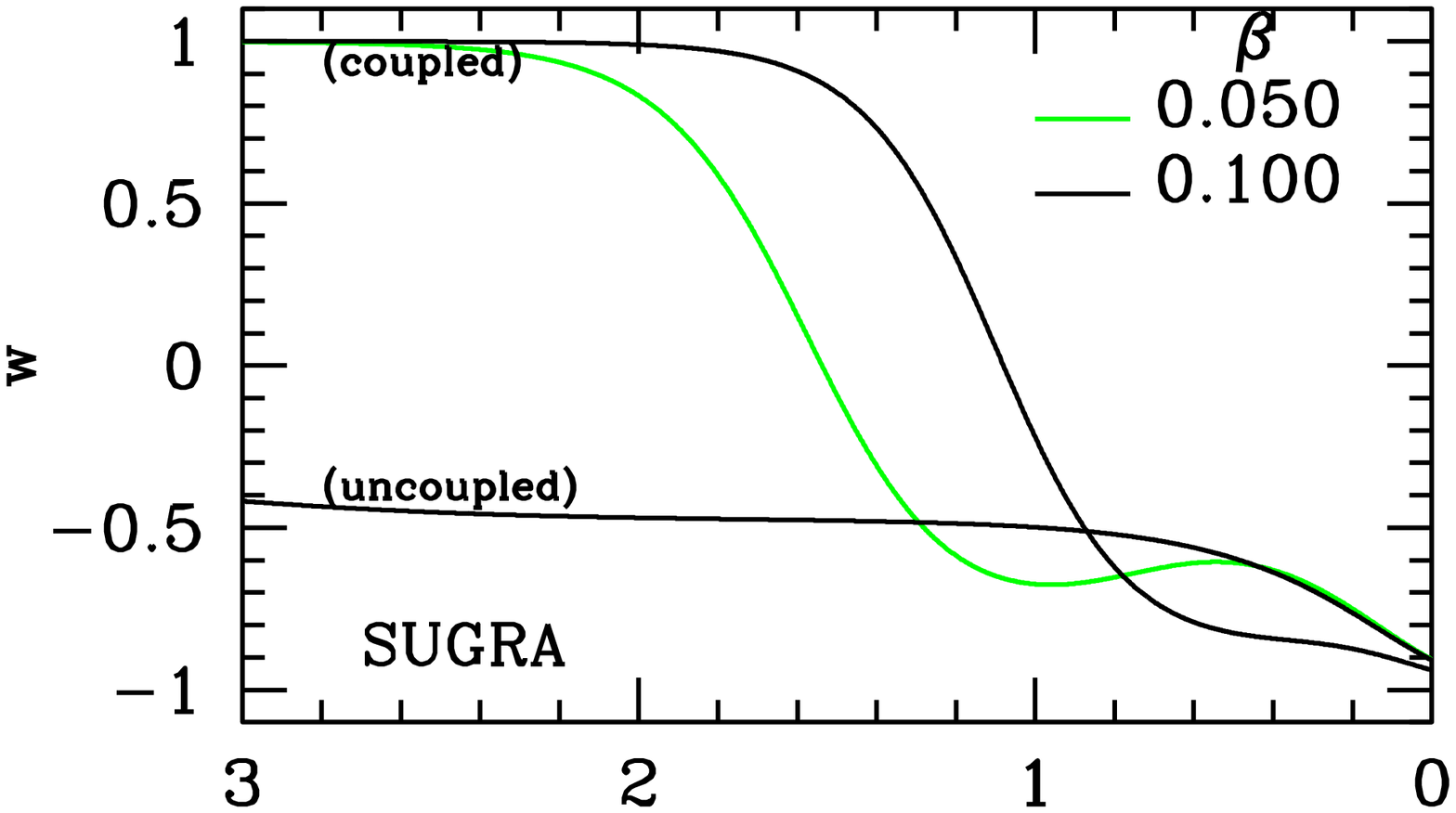}
\vskip -4.6truecm
\includegraphics[scale=.52]{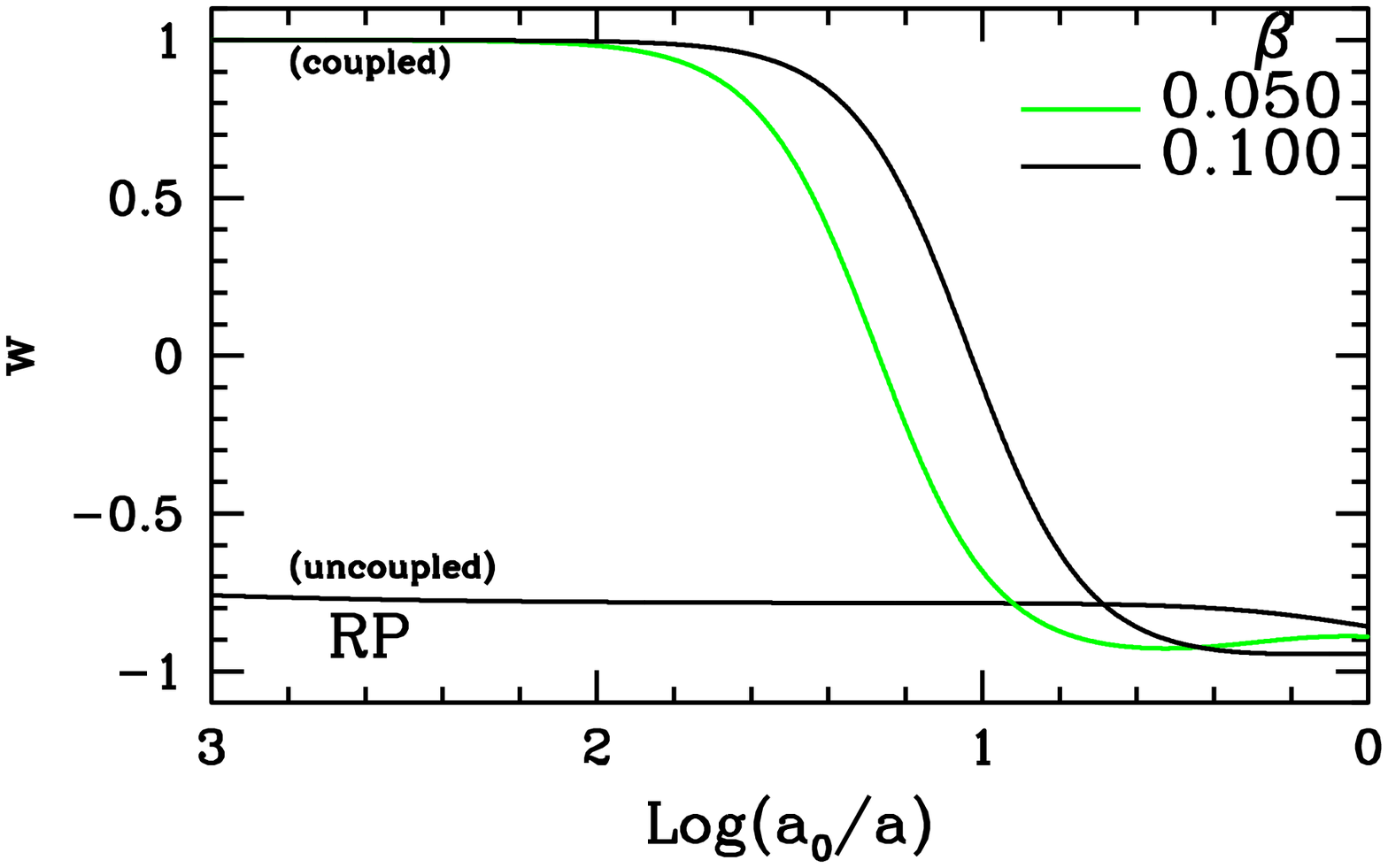}
\end{center}
\vskip -.6truecm
\caption{Scale dependence of the DE state parameter in SUGRA and RP
  models, in the presence and absence of energy transfer from a large
  CDM component. The main parameters of the models are $\Omega_b=
  0.046$, $\Omega_c = 0.209$, $H_0=73~$km/s/Mpc, while $\Lambda =
  0.1~$GeV for SUGRA and $\Lambda = 10^{-5}$GeV for RP. The values of
  the coupling $\beta$ are shown in the frames. }
\label{SURP}
\end{figure}
The Figures \ref{SURP} show the shift of $w$ from +1 to $\sim -1$,
with some potentials studied in the literature. Values $\beta <
\sqrt{3}/2$ are however taken in these Figures, while DE is coupled to
all DM.

Our aim here, however, just amounts to showing that DE can achieve a
density consistent with observations while DM$_{cou}$, on the
contrary, yields a negligible contribution to the cosmic budget.  This
unavoidably requires that the energy density of the $\phi$ field
turns from (prevalently) kinetic to (prevalently) potential.

The only caution to be taken is avoiding a too fast $w$ decrease, as
the expression of $\tilde W$ has a contribution from $dw/da$, which
may become dangerously high.
\begin{figure}
\begin{center}
\vskip -.4truecm
\includegraphics[scale=0.5]{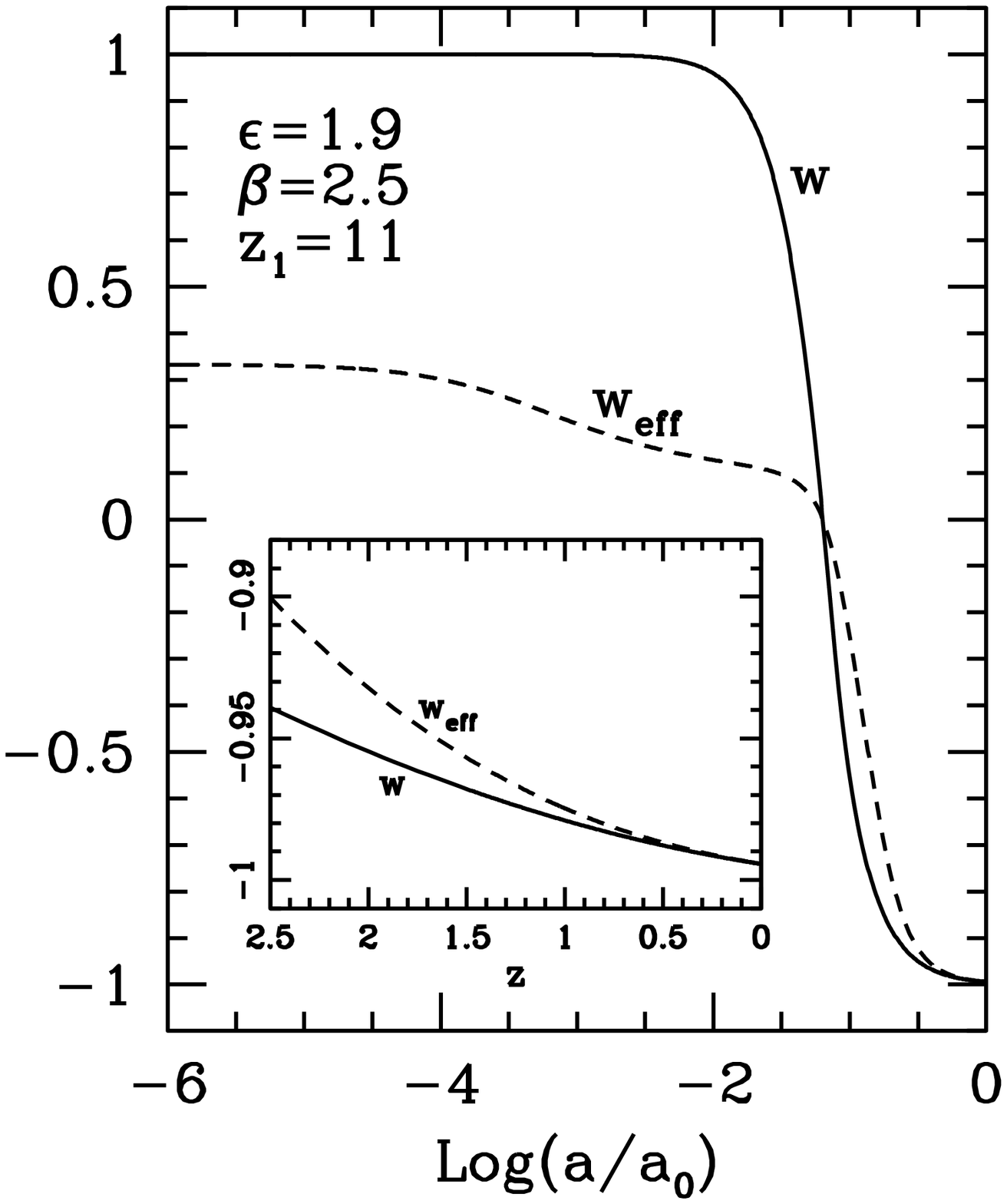}
\end{center}
\vskip -.6truecm
\caption{Solid lines: DE state parameter during the kinetic--potential
  transition. Dashed lines: Effective state parameter, if hypothetical
  data are considered by supposing DE not to be coupled. The large
  $\beta $ value balances the small $DM_{cou}$ density, so that the
  $w$ scale dependence is not so different, e.g., from a RP case with
  $\beta = 0.1$ and all DM coupled (Figure 4). The inner panel show
  the low--$z$ behavior, outlining that $w$ and $w_{eff}$ are not so
  different up to $z \sim 1$, again because of the small $DM_{cou}$
  density parameter.  }
\label{Ustabile}
\end{figure}

In this work we take the following class of interpolatory functions
$$ 
~~~~~~~~~1+w = (1+w_+) \exp\left[-\left(a \over
  a_1\right)^\epsilon\right]~~~~~~{\rm for}~a \leq a_1
$$
\begin{equation}
1+w = {1+w_+ \over e} \left(a_1 \over a \right)^\epsilon~~~~~~~~~~{\rm for}~a>a_1~
\end{equation}
yielding 
$$ 
{a \over 1+w} {dw \over da} = -\epsilon \left(a \over a_1
\right)^\epsilon \leq \epsilon ~~~~~{\rm for}~a \leq a_1
$$
\begin{equation}
\label{interpo}
{a \over 1+w} {dw \over da} = -\epsilon  ~~~~~~~~~~~~~~~~~~~{\rm for}~a>a_1 
\end{equation}
so that the $\tilde W$ correction, in respect to $1+3w$ is $\epsilon$,
at most. Here $w_+$ is the DE equation of state at large $z$.  In
principle, $\epsilon$ is to be fixed so that the DE state equation is
a suitable $ w_-$ at $z=0$. Here we used $\epsilon = 1.9$. In Figure
\ref{Ustabile} we show the resulting $w(a) $ behavior, when $1+z_1 =
1/a_1 = 12~.$

Notice that the $w(a)$ behavior shown in the Figure is not so far from
those obtained from some assigned potentials. In particular, the large
value of $\beta$ is balanced by the low DM$_{cou}$ density which,
however, is not arbitrary, as the initial DM$_{cou}$ density parameter
is dictated by $\beta $ itself.

In Figure \ref{Ustabile} we also plot $w_{eff}$, defined in accordance
with eq.~(\ref{213}). As expected from the required expansion rate, at
large $z$ it is $w_{eff} = 1/3~$. When $w$ falls down and intersects
$w_{eff}$, it also starts to decrease. In the inner frame of the
Figure, the low--$z$ behavior is magnified. Up to $z \sim 1$ the scale
dependence of $w$ and $w_{eff}$ will be distinguishable only through
very refined experiments. Above $z \sim 1$, however, the difference
becomes more relevant.

\begin{figure}
\begin{center}
\vskip -.4truecm
\includegraphics[scale=0.50]{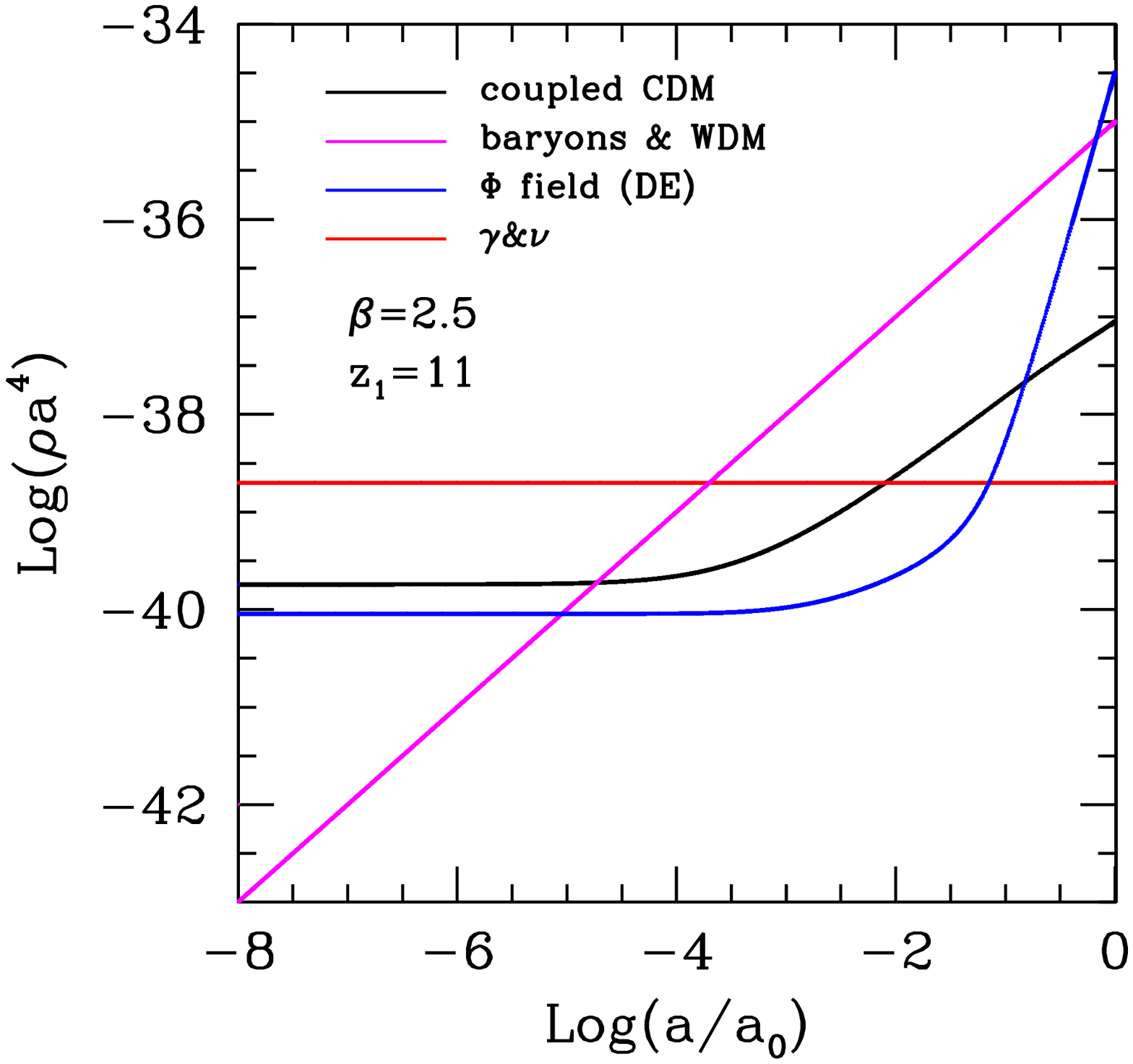}
\end{center}
\vskip -.6truecm
\caption{Density evolutions for a model as in Figure 2, with the DE
  state equation in Figure 5.  }
\label{allc}
\end{figure}
\rm

The interpolation (\ref{interpo}) with $z_1 = 11$ and $\epsilon =
1.9$, then yields the behavior of the different cosmic component shown
in Figure \ref{allc}. Let us specifically outline that the final DE
density can be tuned rather easily to different values; more
specifically, a greater (smaller) $z_1$ yields a greater (smaller)
today's DE density. On the contrary, $DM_{cou}$ final density is
substantially insensitive to the parameter choice.

Altogether, reproducing the observational densities requires a
suitable set of model parameters, but no fine tuning is required:
their tuning must be so precise as the parameter precision required.
In the case of Figure 6, the present DE density is 3 times $DM_{unc}$
(including also baryons), while the final contribution of $DM_{cou}$
to the density budget is in the per mil range.

\section{Astrophysical context: flat galaxy cores and small halo deficit}
If we then compare the cosmological picture proposed here with more
standard scenarios, where DE is however a scalar field, the main
difference is that here the $\phi$ field has substantially contributed
to the cosmic budget since ever.

On the contrary, multiple DM components have been considered by
several authors (see, e.g., \cite{Maccio':2012uh}, \cite{blennow},\cite{baldi}),
for various reasons. In this Section we shall debate the option that
$DM_{unc}$ is WDM, instead of CDM. In association with the presence of
a second DM component, and owing to the peculiar features of
DM$_{cou}$, this option seems particularly appealing.

Let us then first remind that, if a DM component is coupled to DE,
the effective gravity it feels is modified. For baryonic matter, we
have sophisticated tests allowing us to state that, on terrestrial and
planetary scales, no coupling with DE exists \cite{gravity}. No such
test can be extended to DM and this is why quite a few options for
DM--DE coupling were considered in the literature, and tested against
cosmic data.

Observations, however, so nicely fitting most $\Lambda$CDM
predictions, put stringent limits to the coupling $\beta$
\cite{limits}. They can be eased when DM--DE coupling is considered in
association with non--vanishing neutrino masses \cite{mmc}, and Mildly
Mixed Coupled cosmologies were found to fit data slightly better than
$\Lambda$CDM. The likelihood improvement, however, is not
statistically significant and, even within this approach, a coupling
range including $\beta > 0.886$ is excluded. If two DM components
exist, however, we are in a fully different context, that we may
tentatively exploit to seek a solution to some inconsistencies of the
$\Lambda$CDM model, on sub-galactic scales, put in evidence by N--body
simulations, namely if DM is assumed to be ``cold''.

A first difficulty concerns the amount of substructure in Milky Way
sized haloes \cite{klypin-moore}. Models involving CDM overpredict
their abundance by approximately one order of magnitude. A second
issue concerns the density profiles of CDM haloes in simulations,
exhibiting a cuspy behavior \cite{moore1994,
  FloresPrimack1994-Diemand2005-Maccio2007-Springel2008}, while the
density profiles inferred from rotation curves suggest a core like
structure \cite{deBlok2001-KuziodeNaray2009-Oh2011}. A third issue
concerns dwarf galaxies in large voids: recent studies
\cite{Tikhonov2009-Zavala2009-PeeblesNusser2010} re-emphasized that
they are overabundant.

It is known that replacing CDM with a ``warmer'' DM component, as a
thermal relic of particles whose mass is $\sim 2$--3 keV, yields
better predictions. What is essential, however, is the streaming
length of such component. Accordingly, it can also be replaced by
particles of different mass, with a non--thermal distribution, but
similar average velocity. However, there is a number of ``thermal''
candidates for such warm dark matter (WDM); among them, a sterile
neutrino and a gravitino \cite{AbazajianKoushiappas2006-Boyarsky2009a}
find a reasonable motivation in particle theory
\cite{DodelsonWidrow1994-Buchmueller2007-TakayamaYamaguchi2000}.

The long streaming length of such particles causes a strong
suppression of the power spectrum on galactic and sub-galactic scales
\cite{Bond1980-BonomettoValdarnini1984-DodelsonWidrow1994-HoganDalcanton2000-ZentnerBullock2003-Viel2005-Abazajian2006}
and solves several above problems. In particular, the profiles of WDM
haloes, similar to CDM haloes in the outer regions, flatten towards a
constant value in the inner regions, as predicted in
\cite{Villaescusa-NavarroDalal2011} and found in simulations
\cite{Colin2008-Maccio2012}.

However, the core size found is 30--50 pc, while the observed cores in
dwarf galaxies are around the 1000 pc scale
\cite{WalkerPenarrubia2011-JardelGebhard2012}. A dwarf galaxy core in
this scale range would be produced by higher velocity particles, as
those belonging to a thermal distribution if their mass is $<0.1$--0.3
keV. Increasing the velocity, however, yields a greater streaming
length, exceeding the size of these very dwarf galaxies, in the first
place \cite{MaccioFontanot2010}.

In view of these difficulties, the idea that WDM is accompanied by a
smaller amount of CDM has already been put forward
\cite{Maccio':2012uh}. The WDM particle velocities could then be
greater, while a low--mass population is however produced by CDM
clustering. This suggestion was been put forward quite independently
of any particle or cosmic model. In particular, assuming {\it ad hoc}
a twofold dark matter component does not ease coincidence problems.

It is then clear that the model discussed in this paper could been
adapted to meet the above requirement. DM$_{unc}$ would be a kind of
WDM, made of high--speed particles. DM$_{cou}$, then, would be
responsible to create condensation sites on scales smaller than the
DM$_{unc}$ streaming length, after its derelativisation. Its role is
similar to the CDM role in $\Lambda$CDM models, after recombination,
when CDM fluctuations cause baryon accretion on scales where primary
baryon fluctuations had been erased during recombination.

At large scale we then expect a standard primeval fluctuation
spectrum, suitably balanced between $DM_{cou}$ and $DM_{unc}$,
although today's $DM_{cou}$ contribution could be not so significant.
Below the streaming length of $DM_{unc}$, however, only $DM_{cou}$
fluctuation initially remain, to create the seeds for the low scale
fluctuation spectrum, after $DM_{unc}$ derelativization. The amplitude
of the overall DM spectrum can then be expected to be a few times
smaller below the WDM (DM$_{unc}$) streaming length.

These qualitative considerations require a detailed quantitative
confirm. It is not unreasonable, however, that ``secondary'' $DM_{unc}$
fluctuations, although still yielding some low--mass structure,
generate less low--mass haloes than a standard CDM spectrum. The
evolution of the fluctuations in $DM_{cou}$ needs however a direct
inspection. It is known that, at the Newtonian level, its
gravitational self--interaction force is enhanced by a factor
$1+4\beta^2/3~~(\sim 9$ for $\beta \sim 2.5$).  Therefore, after
entering the non--linear regime, $DM_{cou}$ fluctuations, more rapidly
than fluctuations in other components, could evolve into collapsed
objects; their expected features are not easily predictable without a
detailed analysis and should be then compared with observations.

Analogous comments can be made for the size of the cores in low mass
galaxies. Being mostly made of low--mass $WDM$ they can be as large as
required, while the galaxy population does exist thanks to the $DM_{cou}$
spectral seeds.

To put these expectations in a quantitative form we need to study
fluctuation evolution in detail. 

\section{Discussion}
In this work we considered coupled DE theories, when the coupling
constant $\beta$ is large. The first finding is then that a dual
component, made of non--relativistic particles and a scalar field, can
be in equilibrium with the radiative components in the radiative
era. The density parameters of the dual components have then a fixed
ratio
\begin{equation}
\label{c2c}
\Omega_c/\Omega_d = 2,~~~ {\rm while}~~~ (\Omega_c+\Omega_d)\beta^2 = 3/4
\end{equation} 
(if the scalar field energy is prevalently kinetic) and such density
parameters keep constant, as both dual components dilute $\propto
a^{-4}$ as the Universe expands. We dubbed them DM$_{cou}$ and DE
although, at this stage, there is no evidence of the latter being
related to observational DE. Another important finding is that the
dual component is stable: if we set initial conditions violating
(\ref{c2c}), the densities of DM$_{cou}$ and DE change and the
condition (\ref{c2c}) is restored.

The dual component gives place to a natural form of DR. Here we find
that
\begin{equation}
\Delta N_{off} \simeq {5.55 \over \beta^2 - 0.75}~,
\end{equation}
so that $\beta = 2.5$ yields $\Delta N_{off} \simeq 1$. All plots in
this paper are given for this case.

When the expansion regime ceases to be radiative, as the density of a
different non--relativistic component ($\propto a^{-3}$) overcomes the
radiative component, also the densities of DM$_{cou}$ and DE start to
increase. This further DM component is dubbed DM$_{unc}$. DM$_{cou}$
and DE densities, although increasing, however keep well below
DM$_{unc}$, unless the energy of the scalar field shifts from kinetic
to potential.

Let us then recall again that, when we try to fit background data to a
model where DE is coupled to the whole DM and self--interacts through
a standard potential (e.g., Ratra--Peebles or SUGRA), we find that the
$\phi$ field energy shifts from kinetic to potential at a redshift
$z_1 \sim 10 $--20. This is the range where the transition must occur,
also in this case. The example shown in the Figures is for $z_1 = 11$,
but similar proportions of DE are obtained also by slightly shifting
$z_1$ and the parameter $\epsilon$, simultaneously. Large shifts of
such parameters are however not allowed and the epoch of the field
transition from kinetic to potential is well constrained, quite
independently of the freedom we still have to modify its detailed
dynamics.

Let us finally stress that we describe a stationary high--$z$
situation, holding since the decoupling of the DM$_{cou}$ component
from the other particles and, possibly, even before this stage. The
dynamics discussed here could originate from a phenomenological
Lagrangian coupling/mass term
\begin{equation}
{\cal L} = \Gamma \dot \phi \bar \psi \psi~~~
{\rm or} ~~~~
{\cal L} = {\mu\, T \over m_p} \bar \psi \psi~,
\end{equation}
holding since then. Here $\psi$ describes spinor particles, which are
DM$_{cou}$. Should the validity of this Lagrangian extend until the
end of the inflationary era, one wonders whether any relation exist
between the DE $\phi$--field and the scalar field responsible for the
inflationary process itself.

\vskip .4truecm
\noindent
ACKNOWLEDGMENTS - We thanks Matteo Viel, Marino Mezzetti and Luca
Amendola for useful discussions.  S.A.B. acknowledges the support of
CIFS though the contract n.~24/2010 and its extension
Prot.n.2011/338bis~.

\vskip 2.truecm

\end{document}